\documentstyle[12pt]{article}

\let\ssection=\section
\renewcommand{\section}{\setcounter{equation}{0}\ssection}
\newfam\eusmfam
\newfam\scrfam
\batchmode\font\tenscr=rsfs10 scaled \magstep1 \errorstopmode
\ifx\tenscr\nullfont
\else	\font\sevenscr=rsfs7 scaled \magstep1
	\font\fivescr=rsfs5 scaled \magstep1
	\skewchar\tenscr='177 \skewchar\sevenscr='177 \skewchar\fivescr='177
	\textfont\scrfam=\tenscr \scriptfont\scrfam=\sevenscr
	\scriptscriptfont\scrfam=\fivescr
	\def\scr{\fam\scrfam}
	\def\cal{\scr}
\fi

\newcommand{\eqn}[1]{\label{#1}}
\newcommand{\eq}[1]{\begin{equation}#1\end{equation}}

\newcommand{\eqs}[1]{\begin{eqnarray}#1\end{eqnarray}}

\newcommand{\br}[1]{\left(#1\right)}

\def\lb{\nonumber\\}
\newcommand{\refbr}[1]{(\ref{#1})}

\def\a{\alpha}
\def\b{\beta}
\def\g{\gamma}
\def\d{\delta}
\def\e{\epsilon}
\def\f{\phi}
\def\F{\Phi}
\def\z{\zeta}

\def\l{\lambda}
\def\m{\mu}
\def\n{\nu}
\def\r{\rho}

\def\s{\sigma}
\def\S{\Sigma}

\def\t{\tau}
\def\o{\omega}

\def\O{\Omega}
\def\L{\Lambda}
\def\/{\over}
\def\*{\partial}
\def\|{\mid}

\def\2{\half}
\def\3{{1\/3}}
\def\w{\wedge}

\renewcommand{\baselinestretch}{1.5}
\newcommand{\NP}[1]{Nucl. Phys.\ {\bf #1}\ }
\newcommand{\PL}[1]{Phys. Lett.\ {\bf #1}\ }
\newcommand{\CMP}[1]{Comm. Math. Phys.\ {\bf #1}\ }

\newcommand{\CQG}[1]{Class. Quantum Grav.\ {\bf #1}\ }
\newcommand{\PR}[1]{Phys. Rev.\ {\bf #1}\ }
\newcommand{\PRL}[1]{Phys. Rev. Lett.\ {\bf #1}\ }

\newcommand{\IJMP}[1]{Int. J. Mod. Phys.\ {\bf #1} \ }
\newcommand{\LMP}[1]{Lett. Math. Phys.\ {\bf #1} \ }
\newcommand{\half}{\mbox{\scriptsize $ \frac{1}{2}$}}

\newcommand{\LL}{{\cal L}}
\newcommand{\LLO}{{\cal L}_0}
\newcommand{\LLh}{\hat{\cal L}}
\newcommand{\ED}{{\rm Ext}\;D({{\cal L}_0})}
\newcommand{\tr}{{\rm tr}}
\newcommand{\DD}{{\cal D}}
\newcommand{\Dcov}{\DD}
\newcommand{\FF}{{\cal F}}
\newcommand{\HH}{{\cal H}}
\newcommand{\GG}{{\cal G}}
\newcommand{\Tt}{\tilde{T}}

\begin{document}

\newpage
\vspace*{-3cm}
\pagenumbering{arabic}
\begin{flushleft}
G\"{o}teborg\\
ITP 93-37\\
hep-th/9401027\\
December 1993
\end{flushleft}
\vspace{12mm}
\begin{center}
{\huge Higher-dimensional loop algebras, non-abelian extensions and
$p$-branes}\\[14mm]
\renewcommand{\baselinestretch}{1.2}
\renewcommand{\footnotesep}{10pt}
{\large Martin Cederwall\footnote{TFEMC@FY.CHALMERS.SE}\\
Gabriele Ferretti\footnote{FERRETTI@FY.CHALMERS.SE}\\
Bengt E.W. Nilsson\footnote{TFEBN@FY.CHALMERS.SE}\\
Anders Westerberg\footnote{TFEAWG@FY.CHALMERS.SE}\\
}
\vspace{12mm}
{\sl Institute of Theoretical Physics\\Chalmers University of Technology\\ and
University of G\"{o}teborg\\
S-412 96 G\"{o}teborg, Sweden}
\end{center}
\vspace{3mm}
\begin{abstract}
We postulate a new type of operator algebra with a non-abelian extension. This
algebra generalizes the Kac--Moody algebra in string theory and
the Mickelsson--Faddeev algebra in three dimensions to higher-dimensional
extended objects ($p$-branes). We then construct new BRST
operators, covariant derivatives and curvature tensors in the
higher-dimensional generalization of loop space.
\end{abstract}
\renewcommand{\baselinestretch}{1.5}

\newpage

\section{Introduction}

Infinite-dimensional Lie algebras are of crucial importance in a number
of physical applications and have been a subject of intensive study
for a number of years. The most well-known
examples to physicists are the Virasoro \cite{Vir70} and Kac--Moody (KM)
algebras
\cite{Kac1,Moo1,BH1}, which play fundamental roles
in quantum field theories in two dimensions with
applications in statistical physics as well as in string theory and
2d quantum gravity  \cite{ISZ}. Also, in three dimensions one finds the
Mickelsson--Faddeev (MF) algebra \cite{Mic83,Mic85,Fad84},
which arises in connection with
gauge theories interacting with chiral matter (see e.g. \cite{Mic89,Jack} and
refs. therein).

In most situations it is convenient to interpret these infinite-dimensional
algebras as extensions of simpler (classical) algebras of
observables. Roughly speaking, this means that the commutation relations
get augmented by an extra term; a central extension
in the case of the Virasoro and KM algebras, and an abelian
extension in the MF case. These extensions are of profound importance for the
algebras and their representation theories, and, consequently, also for their
physical applications.

In all these cases the additional operators added to the algebra commute among
themselves, hence the name abelian extension.
 In the Virasoro and KM cases they even commute with all other observables, and
are therefore referred to as
central. In this paper we will argue that non-abelian extensions may arise in
the context of the theory of extended objects
 ($p$-branes) coupled to Yang--Mills fields.
We will postulate a new algebra on the $p$-brane that reduces to the KM algebra
for $p=1$ (i.e. the string) and to the MF algebra for $p=3$.
For higher $p$'s our construction gives rise to a
non-abelian extension of the algebra of charge densities on
 the $p$-brane. This is of potential
interest for particle physics because it has been speculated that the $p=5$
case is relevant if one wants to incorporate
non-perturbative effects in superstring theory by appealing to a dual
formulation in terms of the five-brane \cite{FILQ}. Such a duality was
conjectured in \cite{Du88} before the five-brane was proven to exist
\cite{Str90}. The consequences of the existence of such a Montonen-Olive
\cite{MO,GNO,Ol81,Osb79} type duality in string theory might be physically
extremely important \cite{FILQ}, involving phenomena like
supersymmetry breaking and
the occurrence of a non-zero potential for the dilaton.

Recently \cite{DD93,Dix93,PSez92,MP93,BPSST} an attempt has been made to use a
straightforward generalization of the MF algebra to higher
 dimensions in the context of the $p$-branes. In this attempt, the algebra
reads \cite{Dix93}
\eq{ [\tilde T^a(\s), \tilde T^b(\s^\prime)] = if^{abc} \tilde
T^c(\s)\d^p(\s-\s^\prime) + n\e^{i_1\cdots i_p}\partial_{i_1}
N^{(ab)}_{i_2\cdots i_{p-1}}(\s)\partial_{i_p}\d^p(\s-\s^\prime), \eqn{MFFIVE}}
where, in general, the extension $N$ is a function of the Maurer--Cartan form
$K$ on the group $G$ pulled back to the $p$-brane.
The $\tilde T$'s generate gauge transformations under which
$K$ transforms as a gauge field. For $p=1$ \cite{BDS91}, $N$ is independent of
$K$ and the algebra in \refbr{MFFIVE} reduces to the Kac--Moody algebra, while
for
$p=3$ it is linear in $K$ and one obtains the linear algebra in $\tilde T$ and
$K$
described in \cite{DDS,DD93}. This is the Mickelsson--Faddeev
algebra
familiar in the context of chiral gauge theories. For $p>3$ the algebra becomes
non-linear because $K$, as briefly explained in section 2,
 now appears non-linearly in $N$.

In this paper we impose instead linearity and diffeomorphism
invariance to obtain an alternative generalization that effectively extends
the operator content of the theory with the introduction of new forms of
intermediate degree satisfying non-trivial commutation relations among
themselves, i.e., corresponding to a non-abelian extension.
Thus, when generalizing many of the results known from the loop
space formulation of string theory, our algebra makes
 it possible to avoid non-linearities
even for $p>3$. More specifically, we will construct a BRST operator
for the $p$-brane wave functional, a covariant derivative in ``$p$-loop space''
and its related curvature tensor.
To allow for a direct comparison with \refbr{MFFIVE}, we present here the
explicit form of the algebra that we will obtain for $p=5$:
\eqs{
{[}\Tt^a(\s),\Tt^b(\s')]&=&if^{abc}\Tt^c(\s)\d^5(\s-\s')+
	\tilde{k}d^{abc}\*_i\Tt^{c\,ij}(\s)\*_j\d^5(\s-\s'),\lb
{[}\Tt^a(\s),\Tt^{b\,mn}(\s')]&=&if^{abc}\Tt^{c\,mn}(\s)\d^5(\s-\s')-
	\tilde{k}d^{abc}\e^{ijkmn}\*_iT^c_j(\s)\*_k\d^5(\s-\s'),\lb
{[}\Tt^a(\s),T^b_i(\s')]&=&if^{abc}T^c_i(\s)\d^5(\s-\s')+
	k\d^{ab}\*_i\d^5(\s-\s'),\lb
{[}\Tt^{a\,kl}(\s),\Tt^{b\,mn}(\s')]&=&if^{abc}\e^{iklmn}T^c_i(\s)\d^5(\s-\s')+
	k\d^{ab}\e^{iklmn}\*_i\d^5(\s-\s'),\lb
{[}\Tt^{a\,mn}(\s),T^b_i(\s')]&=&0,\lb
{[}T^a_i(\s),T^b_j(\s')]&=&0.
\eqn{OURFIVE}}
The precise meaning of all the symbols appearing in \refbr{OURFIVE} will be
explained in section 3.
It is important to note that $K$ plays no role in this algebra which therefore
is more general than \refbr{MFFIVE}, since $K$ itself
is connected to a special realization based on functional derivatives on the
group manifold $G$. In fact, when constructing \refbr{OURFIVE}
in section 3 no explicit reference is made to $G$ apart from the fact that we
use forms valued in its Lie algebra.
We should also remark at this point that although we construct most quantities
of relevance
for the $p$-brane algebra \refbr{OURFIVE}, we do not present a $p$-brane action
related to it. For the algebra \refbr{MFFIVE}, on the
other hand, as will be discussed in section 2, such an action exists and does
indeed require the Maurer--Cartan form $K$ for its construction.

To our knowledge this is the first time non-abelian extensions appear in
physics. In this paper we will limit our considerations  to
$p$-branes, although our construction also incorporates (for $p=2$) certain
current algebras arising in the canonical formulation
of (2+1)-dimensional non-linear $\s$-models \cite{RSY84,FR92,Fer92}.
We hope to return to further applications in the future.

To conclude this introduction, let us briefly mention the question of
representation theory.
Finding unitary representations of the Mickelsson--Faddeev algebra
has turned out to be an extremely difficult problem.
However, recently there has been some progress along these lines
in the (3+1)-dimensional case
using regularization techniques based on the calculus of
pseudodifferential operators \cite{Mic93}.
In this context, we believe that our algebra,
being linear and not involving directly a gauge potential, could stand a
better chance of having a tractable representation theory. Also, the fact
that there are now two different higher-dimensional current algebras might
be of some help in understanding the difficulties that one encounters
in the search for unitary representations.

The plan of the paper is as follows. In section 2 we review some of the
previous developments emphasizing the role
played by the Maurer--Cartan one-form appearing in the $\s$-model action, the
BRST operators and the covariant derivatives.
This rather detailed account, which also explains the origin of the algebra
\refbr{MFFIVE},
is included in order to facilitate the comparison to our construction which
is first explained for the BRST operator in section 3, and then for the
covariant derivative in section
4. In section 4 we also discuss the corresponding field strength and its
Bianchi identity.
Section 5 is devoted to putting our new algebra into a mathematical
context, explaining how it fits into the theory of extensions in terms of exact
sequences etc. Finally, in section
6 we summarize our findings and make some additional comments. Formulae for
some of the quantities appearing in the descent equations are collected in
appendix A, together with the explicit expressions for the BRST operators
and covariant derivatives.

\section{The origin of the Mickelsson--Faddeev algebra for the $p$-brane}

We  consider the bosonic $p$-brane as being defined by an embedding of the
$(p+1)$-dimensional worldvolume swept out
by the manifold $\S_p$ moving in spacetime $M$, $\S_p$ being the
$p$-dimensional manifold describing the $p$-brane itself. We will restrict
ourselves to local considerations, and
at the present level of understanding we also have to content ourselves with
coupling
the $p$-brane to background fields in spacetime. In particular, we assume the
existence of a background Yang--Mills field $A$ with
gauge group $G$, and of a background antisymmetric $(p+1)$-tensor $B_{p+1}$
that couple to the $(p+1)$-dimensional worldvolume.

The dynamics of the system will be described by introducing a $p$-brane wave
functional $\Phi(x)$ with $x: \S_p \to M$. Our aim
is to describe the\ ``loop space'' functional algebra of operators acting on
$\Phi$, with special emphasis on the BRST operator
$\d$  and the related covariant derivative $\Dcov$. Throughout the paper we use
the convention that $\d$ {\it commutes} with the
exterior derivative $d$. The BRST operator then acts on the
Yang--Mills field $A$ and its ghost $\o$ as
\eqs{&&\d A= d\o + A\o - \o A, \eqn{/dBA}\\
&& \d \o = -\o ^2 \eqn{/dB/o}.}
For the antisymmetric tensor field $B_{p+1}$ and its sequence of ghosts,
denoted by $\L^q_{p+1-q}$ ($q=1,...p+1$ is the ghost number),
one has
\eqs{\d\!B_{p+1} = n\,\o^1_{p+1} -  d\L^1_p, \eqn{/dBB}\\
\d\!\L^q_{p+1-q} = n\,\o^{q+1}_{p+1-q} -  d\L^{q+1}_{p-q} \eqn{/dB/L}.}
(This notation differs from the one used in \cite{DD93}; here the ghosts
are all denoted by $\L$.)
Furthermore, the descent equations for the Chern--Simons form $\o^0_{p+2}$ and
its descendants take the form
$\d\o^q_{p+2-q}=d\o^{q+1}_{p+1-q}\ (q=0,...p+1)$.
The nilpotency of $\d$ on the background fields $B$ and $\L$ above follows
straightforwardly from these equations.

We must now define how $\d$ acts on the $p$-brane wave functional. To this end,
consider first the case $p=1$ (i.e.\ the string).
When acting on the string functional $\F$ we write this operator as
($\S_1\equiv S^1$)
\eq{\d\F =\Bigg( \int_{S^1}[\tr(-\o T_1)+ \L^1_1]\Bigg)\F, \eqn{/d1}}
where the one-form $T_1=\t^a\Tt^a(\s)d\s$ is a Lie algebra-valued operator on
$S^1$. The Lie algebra generators $\t^a$ are assumed to
be hermitian, satisfying $[\t^a,\t^b]=if^{abc}\t^c$ and
$\tr\,\t^a\t^b=\d^{ab}$.
We also assume the existence of a totally symmetric
tensor $d^{abc}\equiv\tr(\{\t^a,\t^b\}\t^c)$. Furthermore, the scalar density
$\Tt^a(\s)$, the
dual of $T_1$, is required to satisfy a KM algebra as a consequence of imposing
$\d^2\F=0$ \cite{BDS91}. Let us also point out that
throughout this paper the pull-back of all the background forms on the
$p$-brane is always understood. For example, in the above
mentioned case we have
\eq{\o=\o(x(\s)),\ \ \ \ \ \ \Lambda_1^1 =
\Lambda_\mu(x(\s)){{dx^\mu}\over{d\s}}d\s,}
$\Lambda_\mu$ being the background form in spacetime.

When we turn to the case $p=3$,
the operator $\d$ becomes slightly more complicated, involving a new operator
$T^a_i(\s)$ \cite{DD93} apart from the scalar density $\tilde{T}^b(\s)$
which is now the dual of a three-form
$T_3$. Using the language of forms and pull-backs, the expression for $\d$
given in \cite{DD93} can be written as
\eq{\d  \F = \Bigg(\int_{\S_3}(\tr(-\o T_3 - {1\/2}n[A,d\o]T_1)
+\L^1_3)\Bigg)\F, \eqn{/d3}}
where $T_1=\t^aT^a_i(\s)d\s^i$.
Enforcing $\d^2\F=0$ in this case leads to an MF type algebra for $\tilde{T}$
and $T_i$.

When considering higher-dimensional cases, there are at least two alternative
ways of generalizing the algebras
of operators appearing for $p=1,3$. The first such construction occurring in
the literature \cite{DD93,Dix93} relies upon a specific
construction of the $p$-brane by a Kaluza--Klein approach. As emphasized in the
introduction, this method
introduces certain non-linearities, and the object of the rest of this section
is to explain how this comes about.
In section 3 we will then present a new, alternative, generalization that makes
use of a linear algebra with a non-abelian
extension.

Let us thus review the Kaluza--Klein approach to the construction of
$p$-branes. This approach relies on the observation
that the operator $T^a_i$ can be identified with the Maurer--Cartan form
$K^a_i$ on the group manifold $G$. One can then
proceed to construct a locally gauge invariant action for the $p$-brane of the
form
\eq{S_{p+1}=S_{p+1}^K+S_{p+1}^W \eqn{S}.}
Here the kinetic part is $(i,j=0,1,..,p)$
\eqs{S_{p+1}^K=\int d^{p+1}\s (-{1\/2}\sqrt{-\g})(\g ^{ij}g_{ij}+\g ^{ij}\tr
J_iJ_j-(p-1)), \eqn{SK}}
with $\s^i,\g^{ij}$ ($\g=det\g_{ij}$) the worldvolume
coordinates and metric, respectively, and $g_{ij}(x)$ the
pull-back to the worldvolume of the
background spacetime metric $g_{\m \n}$ $(\m,\n=0,1,..,D-1)$.
The current $J=d\s ^i J_i$ is expressible
in terms of the pull-backs of the Yang--Mills field $A$ and the Maurer--Cartan
form $K$ on the
group $G$ as $J=A-K$. The last term in \refbr{SK} is the cosmological constant
which is needed to make it possible to solve
for the induced metric $g_{ij}$ by means of its field equations.

The WZW part of the $p$-brane action reads \cite{DDS}
\eq{S^W_{p+1}=\int(B_{p+1}+C_{p+1}(A,K)-b_{p+1}), \eqn{SW}}
where $B$ and $b$,  respectively, are the pull-backs to the worldvolume of the
antisymmetric tensor fields
$B_{\m_1...\m_D}(x)$ in spacetime and $b_{m_1..m_D}(y)$ on the group manifold
$G$.
Finally, the quantity $C_{p+1}(A,K)$ appearing in $S^W$
must transform under gauge transformations in such a way as to ensure that the
action $S$ be gauge invariant.

For the string (i.e. $p=1$) $C_2$ is easily found by recalling that in the low
energy limit of the heterotic string,
supergravity is coupled to super-Yang--Mills \cite{BRWN,ChMo} which
forces one to let $B_2$ transform under gauge transformations according to
\eq{\d_{\l}B_2 = \tr (Ad\l),  \eqn{/dB2}}
where $\l^a$ is an $x$ dependent gauge parameter. This conclusion can also be
seen to follow from a Kaluza--Klein procedure
\cite{DNP85,DNPW86}
in which the bosonic string is compactified on $M_{10}\times G$, where $M_{10}$
is the 10-dimensional Minkowski space and $G$ the group
manifold $E_8\times E_8$ or $SO(32)$. This approach also provides us with the
relation
\eq{h_3 = db_2 + \o^0_3(K)=0. \eqn{h3}}
Here $\o ^0_3(K)$ is the Chern-Simons functional with $A$ replaced by the
Maurer--Cartan form $K$. The explicit form of $\o ^0_3(K)$
and some other quantities appearing in the descent equations can be found in
the appendix.
In fact, by appealing to the descent equations we conclude that choosing
$\d_{\l}b_2=\o ^1_2(K,\l)$ makes
$h_3$ is invariant under gauge transformations.

Since the kinetic part \refbr{SK} of the action \refbr{S} is trivially gauge
invariant for $p=1$, invariance for the complete
action is achieved if
\eq{\d_{\l}C_2 = -\o^1_2(A,\l)+\o^1_2(K,\l). \eqn{/dC}}
To get this result we have also made use of the observation that $\d_{\l}B_2 =
\tr (Ad\l)=\o^1_2(A,\l)$.
Then, by setting $C_2(A,K)=\tr (AK)$ the above expression for $\d_{\l}
C_2(A,K)$ becomes a direct consequence of
the gauge transformation properties $\d_{\l}A=d\l+[A,\l]$ and
$\d_{\l}K=d\l+[K,\l]$.
This also allows us to check the gauge covariance of the current $J=A-K$ in a
trivial manner.

These quantities are easily generalized to arbitrary odd $p$, as explained in
\cite{DDS}, and the action \refbr{S}
can be verified to be gauge invariant in exactly the same way as for $p=1$.
Thus, once the explicit forms of
$\o ^q_{p+2-q}$ for $q=0,1$, and $C_{p+1}(A,K)$ are found, the invariance can
be established. The latter form
\cite{DDS,Dix93} can be found by adopting the technique given in \cite{MSZ} for
deriving solutions to the descent equations.
Explicit formulae for $p=3,5$ can be found in \cite{PSez92,MP93}.

In the above approach, with the identification
$T_i^a=K_i^a$, one obtains for $p>3$ an extension of the commutation relations
of the charge densities by a composite operator $N$ that is non-linear in the
$K$'s:
\eq{ [\tilde T^a(\s), \tilde T^b(\s^\prime)] = if^{abc} \tilde
T^c(\s)\d(\s-\s^\prime) + n\e^{i_1\cdots i_p}\partial_{i_1} N^{(ab)}_{i_2\cdots
i_{p-1}}(\s)\partial_{i_p}
\d(\s-\s^\prime).}

The explicit expression for $N$ can easily be found by considering the
form of the cocycle $\omega^2_p(K,\lambda,\lambda^\prime)$ and
substituting two delta functions for the gauge parameters $\lambda$
and $\lambda^\prime$. For $p=5$ one obtains
\eq{ N^{(ab)}_{ijk}= \omega^{abcd}U^c_{[ij}K^d_{k]}, }
with
\eq{U^c_{ij}= 5(\partial_i K^c_j - \partial_j K^c_i) + 2f^{cef}K_i^e K_j^f}
and
\eq{\omega^{abcd}=\tr(2\t^a\t^b\t^c\t^d + 2\t^b\t^a\t^c\t^d +
\t^a\t^d\t^b\t^c + \t^a\t^c\t^b\t^d).}

Since the $K$'s commute among themselves, $N(\s)$ also
has vanishing commutators with $N(\s^\prime)$ and the $K$'s. In the next
section we will show that,
by enlarging the algebra of operators in loop space, it is possible to obtain a
linear algebra from which all the
relevant functional operators in loop space can be constructed in a natural
way. The form of this new algebra is fixed uniquely by the requirements of
linearity and diffeomorphism invariance on
$\S_p$.

\section{Higher-dimensional loop algebras with non-abelian extensions and the
construction of the $p$-brane BRST operator}

In this section we will introduce our new algebra in its original and most
general form.
The first step in this process is to rewrite the action of the BRST operator on
the string and three-brane wave functionals, given in
\refbr{/d1} and \refbr{/d3}, respectively, in a unified and compact way that
immediately generalizes to any odd $p$:
\eq{\d\F =\br{T(R)+\int_{\S_p}\L^1_p}\F. \eqn{/dp}}
Here $R\,$ is a formal sum of Lie algebra-valued forms in spacetime of even
degree pulled back to $\S_p$, constructed from the
Yang--Mills field $A$ and its ghost $\o$. Furthermore, $T$ is a linear
functional of $R$ defined by
\eq{T(R) = \int_{\S_p} \tr\,TR,\eqn{TRdef}}
where $T$ on the right hand side is a formal sum of operator-valued forms on
$\S_p$ of odd degree. The integral is assumed to
vanish on all forms of degree different from $p$. We can therefore restrict the
formal sums to $R=R_0+R_2+\cdots+R_{p-1}$ and
$T=T_p+T_{p-2}+\cdots+T_1$, respectively, and so
\eq{T(R) = \sum_{0\leq q \leq (p-1)/2} \int_{\S_p}\tr\,T_{p-2q}R_{2q}.}
Note also that, since the BRST operator raises the ghost number by one, the
Yang--Mills ghost $\o$ must appear linearly in $R$.

In particular, for $p=1$, the expression \refbr{/d1} for $\d\F$ from section 2
is recovered by choosing $R\equiv R_0=-\o$ and
$T\equiv T_1=\t^a\Tt^a(\s)d\s$, $\Tt^a(\s)$ being a Kac--Moody algebra
generator as a consequence of imposing $\d^2\F=0$.
Similarly, for $p=3$ we must choose $R\equiv R_0+R_2=-\o-{1\/2}n[A,d\o]$ and
$T\equiv T_3+T_1$, with $T_3=\t^a\Tt^a(\s)d^3\s$ and
$T_1=\t^a T^a_i(\s)d\s^i$ in this case forming a Mickelsson--Faddeev algebra
due to the nilpotency requirement.

Thus, in section 2 the commutation relations for the operators $T_p$,
$T_{p-2}$,...,$T_1$ were considered to follow by enforcing
nilpotency of the BRST operator. The idea is now to turn this argument around,
and consider instead the {\it algebra} as the
fundamental structure. We thus first postulate an algebra for the $T$'s and
then ask for the expression for $R$ that makes it
possible to construct a nilpotent BRST operator. In this process we are led by
the requirements of diffeomorphism invariance,
which, in fact, already has led us to the expression \refbr{/dp}, and,
furthermore, that of agreement with the algebras
previously obtained for $p=1,3$. In order to make the comparisons we will
therefore start by considering these two cases.
Also, when formulating the algebra, it is natural to use, instead of $R$,
formal sums $X=X_0+X_2+\cdots X_{p-1}$ of
Lie algebra-valued forms of ghost number zero.

Hence, assume first that $p=1$. Consider two zero-forms $X$ and $X'$, and a
one-form operator $T$ on $S^1$, all Lie algebra-valued.
Then form $T(X)$ and $T(X')$ as in \refbr{TRdef} and assume that these objects
satisfy the algebra
\eq{[T(X),T(X')]=T([X,X'])+k\int_{S^1}\tr XdX'. \eqn{TT1}}
Taking $X(\s'')=\t^a\d(\s''-\s)$ and $X'(\s'')=\t^b\d(\s''-\s')$, it
immediately follows that this algebra is equivalent to a
Kac--Moody algebra at level $k$:
\eq{[\tilde{T}^{a}(\s),\tilde{T}^{b}(\s
')]=if^{abc}\tilde{T}^{c}(\s)\d(\s-\s')+k\d^{ab}\*_{\s}\d(\s - \s'). \eqn{TTk}}
Thus, $\tilde{T}(\s)$, the dual of the one-form $T_1$, is nothing but the
current in terms of which the Kac--Moody algebra is usually
expressed. Expanded in modes on $S^1$ according to
$\tilde{T}^a(\s)=\sum_nJ^a_ne^{in\s}$, it gives the algebra in its most common
form:
\eq{[J^a_m,J^b_n]=if^{abc}J^c_{m+n}+k\d^{ab}m\d_{m+n,0}. \eqn{JJk}}
(Recall our conventions $[\t^a,\t^b]=if^{abc}\t^c$ and $\tr\,\t^a\t^b=\d^{ab}$
for the Lie algebra of $G$.)
Expressed in terms of the anticommuting variable $R$, the algebra \refbr{TT1}
reads
\eq{\{T(R),T(R')\}=T(\{R,R'\})+k\int_{S^1}\tr R\,d\!R'. \eqn{TT1R}}

Having postulated the algebra, the next step is to impose nilpotency on the
BRST operator when acting on the string wave
functional $\F$. From \refbr{/dp} we find
\eqs{
\d^2\F&=&\d\br{(T(R)+\int_{S^1}\L^1_1)\F} \nopagebreak[4]\lb
	&=&(T(\d R)+\int_{S^1}\d\L^1_1)\F-(T(R)+\int_{S^1}\L^1_1)\d\F
\nopagebreak[4]\lb
	&=&(T(\d R)+\int_{S^1}\d\L^1_1)\F-(T(R)+\int_{S^1}\L^1_1)^2\F, \eqn{delta2}
}
where $\d R$ refers to the BRST variation of the background fields appearing in
$R$.
Using \refbr{delta2}, the algebra \refbr{TT1R} and the descent equation
\refbr{/dB/L}, the nilpotency equation $\d^2\F=0$ then
can be written as two separate equations,
\eqs{\d\!R-R^2=0, \eqn{RR1}\\
\int_{S^1}(n\,\o^2_1 - {1\/2}k\,\tr Rd\!R)=0, \eqn{RR2}}
the latter one corresponding to the central part. It is easily seen that
\refbr{RR1} is satisfied by $R=-\o$, while \refbr{RR2} gives
the additional relation $k=2n$.

In the case of the three-brane, for which $T=T_3+T_1$ and $R=R_0+R_2$, we make
the observation that in the wedge product of
$T$ and $\{d\!R,d\!R'\}$ there appears a term $T_1\{d\!R_0,d\!R_0'\}$ of degree
three, i.e., $T(\{d\!R,d\!R'\})\neq 0$.
Hence, we take the generalization of the algebra \refbr{TT1R} for $p=3$ to
include also terms of this kind:
\eq{\{T(R),T(R')\}=T(\{R,R'\}+\tilde{k}\{d\!R,d\!R'\})+k\int_{\S_3}\tr
R\,d\!R'. \eqn{TT3R}}

Introducing the term multiplying $\tilde{k}$ makes \refbr{TT3R} equivalent to
the MF algebra, as we will now show, by extracting
the explicit form of the algebra. Apart from enabling the comparison this will
reveal the different roles played by the two
parameters $k$ and $\tilde{k}$.
For this purpose, it is convenient to return to the formulation in terms of
commuting arguments $X$ and $X'$, i.e.\,
\eq{[T(X),T(X')]=T([X,X']+\tilde{k}[dX,dX'])+k\int_{\S_3}\tr XdX'. \eqn{TTX3}}
After decomposition this algebra reads
\eqs{
[\Tt^a(\s),\Tt^b(\s')]&=&if^{abc}\Tt^c(\s)\d^3(\s-\s')-
	\tilde{k}d^{abc}\e^{ijk}\*_iT^c_j(\s)\*_k\d^3(\s -\s'), \eqn{TT1X}\\
{[}\Tt^a(\s),T^b_i(\s')]&=&if^{abc}T^c_i(\s)\d^3(\s
-\s')+k\d^{ab}\*_i\d^3(\s-\s'), \eqn{TT2X}\\
{[}T^a_i(\s ),T^b_j(\s')]&=&0, \eqn{TT3X}}
where we have used $X(\s'')=X_0(\s'')+X_2(\s'')$ with
$X_0(\s'')=\t^a\d^3(\s''-\s)$ and
$X_2(\s'')=\linebreak\t^a\d^3(\s''-\s)d\s''^j\w d\s''^k$, and the analogous
expression for $X'(\s'')$ with $a$ replaced by $b$ and $\s$
replaced by $\s'$.
In this notation, the commutation relations above correspond to the ones for
$[T(X_0),T(X_0')]$, $[T(X_0),T(X_2')]$ and $[T(X_2),T(X_2')]$,
respectively.
Note that while the $k$ term in \refbr{TTk} for the string gives rise to the
central extension, the corresponding term
for $p=3$ in \refbr{TTX3} appears instead in the second commutator
\refbr{TT2X}. Furthermore, the extension of the
first commutator \refbr{TT1X} has become operator-valued since it
originates from the new term in \refbr{TT3R}
multiplying $\tilde{k}$. Rescaling by $k$ makes the operator $T^a_i$ transform
as a gauge field and turns the above algebra into
the Mickelsson--Faddeev algebra. Note also that the algebra is referred to as
having an abelian extension due to the last
commutator \refbr{TT3X} above.

With the new term included in $T(X)$ the previously derived conditions become
\eqs{\d\!R-R^2-\tilde{k}\,(d\!R)^2=0, \eqn{RR}\\
\int_{\S_3}(n\o^2_3 - {1\/2}k\,\tr Rd\!R)=0. \eqn{RdR}}
To solve these equations we use the ansatz
\eq{R=-\o+\z_1\,Ad\o+ \z_2\,d\o A. \eqn{R3ab}}
(Terms involving undifferentiated $\o$'s may also be introduced but the
solution will require their coefficients to vanish.)
Inserted in \refbr{RR} and \refbr{RdR} this ansatz gives $\z_1-\z_2 =
\tilde{k}$ and $(\z_1 - \z_2)k + n=0$, respectively.
We thus conclude that $n=- k\tilde{k}$ and that
\eq{R=-\o +\tilde{k}\br{{1\/2}[A,d\o]+(s-{1\/2})\{A,d\o\}}, \eqn{R3t}}
where $s$ is an arbitrary parameter.

Before giving the ansatz for $R\,$ in the case $p=5$, we make the observation
that the algebra \refbr{TTX3} for $T(X)$ (and
hence also \refbr{TT3R} for $T(R)$) generalizes immediately to any odd $p$
without any further additions, and, consequently,
so do the two conditions \refbr{RR} and \refbr{RdR} above.
The number of terms in the ansatz for $R$ increases rapidly, and already for
$p=5$ it is advisable to do the algebra on a computer.
The condition
\eq{\int_{\S_p}(n\o^2_p - {1\/2}k\,\tr Rd\!R)=0 \eqn{RdRp}}
generalizing \refbr{RdR}, may for that purpose be written in the form
\eq{n\d\o^1_{p+1} - {1\/2}k\,\tr(d\!R)^2=0, \eqn{dRdR}}
with the understanding that only terms of form degree $p+1$ should be
considered. Eq.\ \refbr{dRdR} was obtained by requiring the form
$\chi\equiv n\o^2_p - {1\/2}k\,\tr Rd\!R$ inside the integral to be exact, and
then using the descent
equation $d\o^2_p=\d\o^1_{p+1}$. More precisely, \refbr{dRdR} is the
requirement that $\chi$ be closed, i.e.\ that $d\chi=0$.
To be correct one should thus also add the condition $\chi=d\Upsilon$ for some
$(p+1)$-form $\Upsilon$. However, this condition
turns out to be automatically fulfilled by the solutions to \refbr{RR} and
\refbr{RdRp} for the cases at hand.

The most general ansatz (without terms involving undifferentiated ghosts that
would vanish anyway) contains ten parameters,
not including the one present in the solution for $p=3$:
\eqs{R & = &  -\o + \tilde{k}\br{{1\/2}[A,d\o] + (s-{1\/2})\{A,d\o\}} \lb
 & + & \z_1 A^3\,d\o + \z_2 A^2\,d\o\,A + \z_3 A\,d\o\,A^2 + \z_4 d\o\,A^3 \lb
 & + & \z_5 d\!A\,A\,d\o + \z_6 A\,d\!A\,d\o + \z_7 d\!A\,d\o\,A \lb
 & + & \z_8 A\,d\o\,d\!A + \z_9 d\o\,d\!A\,A + \z_{10} d\o\,A\,d\!A.
\eqn{RANS5}}
Subjected to the two nilpotency conditions this ansatz gives the following
four-parameter solution for $R$:
\eqs{
R & = & -\o + \tilde{k}\br{{1\/2}[A,d\o] + (s-{1\/2})\{A,d\o\}} \lb
  & + & \tilde{k}^2\Bigg((-{2\/3}+v) A^3d\o + ({1\/3}+u) A^2d\o\,A + u
A\,d\o\,A^2 + v d\o\,A^3 \lb
      & + & t d\!A A\,d\o + (-1-u+v) A\,d\!A\,d\o + (s+t) d\!A\,d\o\,A \lb
      & + & (-{1\/3}+s^2+t) A\,d\o d\!A + (-u+v) d\o d\!A A + ({2\/3}-s+s^2+t)
d\o\,A\,d\!A\Bigg).
\eqn{R3}}

Having established the existence of a solution, it may be of interest to write
out the explicit form of the algebra.
In order to do that we first note that $T(R)$ now involves three
operator-valued quantities, namely the five-form
$T_5=\t^a\Tt^a(\s)d^5\s$, the three-form $T_3$, with the dual
$\Tt^{a\,ij}(\s)$, and finally the one-form $T_1$.
Repeating the steps described above for the cases $p=1,3$ generates the
following algebraic structure: \pagebreak[3]
\eqs{
{[}\Tt^a(\s),\Tt^b(\s')]&=&if^{abc}\Tt^c(\s)\d^5(\s-\s')+
	\tilde{k}d^{abc}\*_i\Tt^{c\,ij}(\s)\*_j\d^5(\s-\s'),\eqn{TT51X}\\
{[}\Tt^a(\s),\Tt^{b\,mn}(\s')]&=&if^{abc}\Tt^{c\,mn}(\s)\d^5(\s-\s')-
	\tilde{k}d^{abc}\e^{ijkmn}\*_iT^c_j(\s)\*_k\d^5(\s-\s'),\eqn{TT52X}\\
{[}\Tt^a(\s),T^b_i(\s')]&=&if^{abc}T^c_i(\s)\d^5(\s-\s')+
	k\d^{ab}\*_i\d^5(\s-\s'),\eqn{TT53X}\\
{[}\Tt^{a\,kl}(\s),\Tt^{b\,mn}(\s')]&=&if^{abc}\e^{iklmn}T^c_i(\s)\d^5(\s-\s')
	+k\d^{ab}\e^{iklmn}\*_i\d^5(\s-\s'),\eqn{TT54X}\\
{[}\Tt^{a\,mn}(\s),T^b_i(\s')]&=&0,\eqn{TT55X}\\
{[}T^a_i(\s),T^b_j(\s')]&=&0.\eqn{TT56X}}
We can now explicitly see that the algebra of the scalar densities $\Tt^a$ now
has a non-abelian extension; the commutator \refbr{TT51X}
is augmented by an operator $\Tt^{c\,ij}$ that according to \refbr{TT54X} does
not commute with itself. Also, note from \refbr{TT53X}
that the operator $T^a_i$ (properly rescaled by $k$) still transforms as a
gauge field.

\section{Loop space covariant derivatives and non-abelian extensions}

In the previous section we have shown how, given our $p$-brane algebra, one can
construct a nilpotent BRST operator
acting on the $p$-brane wave functional and on the background fields. However,
this is only one of the many ingredients
needed to formulate the higher-dimensional analogue of loop calculus for the
$p$-brane.  We should also specify what we
mean by the connection one-form and the curvature two-form. Actually, from a
logical standpoint, it would have been more
appropriate to begin with the construction of these two objects and then
describe the BRST operator. We have reversed the
order in this paper because the construction of the BRST operator is somewhat
simpler, not having to rely on the definition
of differential forms in higher-dimensional loop space.

It turns out that the extension of ordinary exterior calculus to these spaces
is fairly straightforward, at least for our
purposes. Let us start by choosing a basis in the space of one-forms, which we
denote by $\d x^\mu(\s)$.
In particular, on an arbitrary background vector field
$\Psi = \int d\s^\prime\psi^\nu(x(\s^\prime)){\d\/{\d x^\nu(\s^\prime)}}$
in loop space it takes the value
$\d x^\mu(\s)(\Psi) = \psi^\mu(x(\s))$.

One can use this fact to construct an exterior derivative operator $D$ in loop
space, by setting
$D=\int d\s^\prime\d x^\mu(\s){\d\/{\d x^\mu(\s)}}$, but it is more convenient
never to use the explicit expression and instead rely on the formal
definition of the exterior derivative when doing the calculations. The simplest
case is when the operator $D$ acts on a scalar
function in loop space. To be specific, let us take the scalar $T(R)$, whose
exterior derivative is needed anyway and which can be
regarded as a paradigm for all the following calculations. On such a function,
$D$ acts by creating a loop space one-form, which
can be specified by giving its value on an arbitrary vector field such as
$\Psi$ defined above:
\eq{ DT(R)(\Psi) \equiv \Psi(T(R)) = T(\LL_\psi R). \eqn{DTR}}
Here $\LL_\psi$ represents the Lie derivative with respect to the {\it
spacetime} vector field $\psi^\mu(x)$ and the argument
of $T$ is understood to be pulled back to the $p$-brane after the derivative
has been taken.

The first equality in \refbr{DTR} is simply the definition of the action of the
exterior derivative on a loop space scalar.
The second relation is slightly non-trivial and can be proven by the following
explicit calculations. Instead of being too
general, let us consider a specific example to show how the calculation goes.
The generalization from this example should be straightforward.
Thus, let us consider $T(R)$ of the form
\eqs{T(R)=\int  d\s\e^{ijk}\tr\,T_i(\s)
R_{\mu\nu}(x(\s)){{\* x^\mu(\s)}\over{\* \s^j}}
{{\* x^\nu(\s)}\over{\* \s^k}}. \eqn{TREXAMPLE}}
(This expression arises as one of the factors in the construction of the BRST
operator for the three-brane.)
The second equality in \refbr{DTR} is then obtained by acting on
\refbr{TREXAMPLE} with the vector $\Psi$ defined above:
\eqs{ \Psi(T(R))&=&\int d\s d\s^\prime\e^{ijk}
\tr\, T_i(\s)\psi^\r(x(\s^\prime)){{\d}\over
{\d x^\r(\s^\prime)}}\left(R_{\mu\nu}(x(\s)){{\* x^\mu(\s)}\over{\* \s^j}}
{{\* x^\nu(\s)}\over{\* \s^k}}\right)\lb
&=&\int d\s d\s^\prime\e^{ijk}
\tr\, T_i(\s)\psi^\r(x(\s^\prime))\left(\*_\r R_{\mu\nu}
(x(\s))\d(\s-\s^\prime){{\* x^\mu(\s)}\over{\* \s^j}}
{{\* x^\nu(\s)}\over{\*\s^k}}\right.\lb
&+&\left. R_{\mu\nu}(x(\s))\d_\r^\mu{{\*\d(\s -
\s^\prime)}\over{\*\s^j}}{{\* x^\nu(\s)}\over{\*\s^k}} +
R_{\mu\nu}(x(\s)){{\* x^\mu(\s)}\over{\*\s^j}}
\d_\r^\nu{{\*\d(\s -
\s^\prime)}\over{\*\s^k}}\right) \lb
&=&\int d\s d\s^\prime\e^{ijk}
\tr\, T_i(\s)\psi^\r(x(\s^\prime))\left(\*_\r R_{\mu\nu}
(x(\s))\d(\s-\s^\prime){{\* x^\mu(\s)}\over{\*\s^j}}
{{\* x^\nu(\s)}\over{\*\s^k}}\right.\lb
&-& \left. R_{\mu\nu}(x(\s))\d_\r^\mu{{\*\d(\s -
\s^\prime)}\over{\*\s^{\prime j}}}{{\* x^\nu(\s)}\over{\*\s^k}}
-R_{\mu\nu}(x(\s)){{\* x^\mu(\s)}\over{\*\s^j}}
\d_\r^\nu{{\*\d(\s -
\s^\prime)}\over{\*\s^{\prime k}}}\right)\lb
&=&\int d\s\e^{ijk}
\tr\, T_i(\s)\Bigg(\psi^\r(x(\s))\*_\r R_{\mu\nu}
(x(\s)) \lb  &+& \*_\mu\psi^\r(x(\s))
R_{\r\nu}(x(\s)) +
\*_\nu\psi^\r(x(\s))R_{\mu\r}(x(\s))\Bigg)
{{\* x^\mu(\s)}\over{\*\s^j}}
{{\* x^\nu(\s)}\over{\*\s^k}}\lb
&=&\int d\s\e^{ijk}
\tr\, T_i(\s)(\LL_\psi R)_{\mu\nu}(x(\s))
{{\* x^\mu(\s)}\over{\*\s^j}}
{{\* x^\nu(\s)}\over{\*\s^k}} \ =\ T(\LL_\psi R).\eqn{EXPLICIT}}

The next case of interest is when $D$ acts on a one-form $\xi$ in loop space.
Again, from the general expressions for the exterior derivative one has
\eq{D\xi(\Psi_1,\Psi_2) = \Psi_1(\xi(\Psi_2)) - \Psi_2(\xi(\Psi_1)) -
\xi([\Psi_1,\Psi_2]), \eqn{DONEFORM}}
when the resulting two-form is evaluated on two arbitrary vectors $\Psi_1$ and
$\Psi_2$ in loop space. Using the
well-known relation $[\LL_{\Psi_1}, \LL_{\Psi_2}] = \LL_{[\Psi_1,\Psi_2]}$
for the Lie derivative, one can check directly that $D$ is nilpotent, i.e., for
example, $D^2 T(R)=0$.

The expression for a generic $n$-form $\xi_n$ is well-known from elementary
differential geometry and it is given
here for completeness:
\eqs{ & & D\xi_n(\Psi_1,\cdots,\Psi_{n+1}) \ =  \lb
     &=& \sum_{1\leq i < j \leq
n+1}(-1)^{i+j}\xi_n([\Psi_i,\Psi_j],\Psi_1,\cdots,\Psi_{i-1},\Psi_{i+1},\cdots
\Psi_{j-1},\Psi_{j+1},\cdots,\Psi_{n+1})\lb
&+&\sum_{1\leq i \leq n+1}(-1)^{i+1}\Psi_i(
\xi_n(\Psi_1,\cdots,\Psi_{i-1},\Psi_{i+1},\cdots,\Psi_{n+1})).
\eqn{GENERALD}}
(The only additional case we will need in this paper is $n=2$ for the Bianchi
identities
of the curvature tensor.)
Now that we have the expression for the exterior derivative, we can look for a
covariant version of it by adding
to $D$ a connection one-form. The idea is the same as for the construction of
the BRST operator; we start with an
ansatz in terms of background forms in spacetime. Such an ansatz must be
compatible with our algebra and it must be
diffeomorphism invariant. We then impose the covariance of the derivative in
loop space and obtain the differential
equations that the background fields must obey in spacetime.

Again, we find it very convenient to work with formal sums of differential
forms of various degree. Thus, let $\O_\psi$ be a formal sum of forms of even
degree and $C_\psi$ a $p$-form, both pulled back to the
$p$-brane manifold $\S_p$ and depending on
the background Yang--Mills field $A(x)$ and an arbitrary background vector
$\psi(x)$. (The dependence on the vector appears
through the contraction of one of the indices by $\psi^\mu$ before the
pull-back. Explicit expressions for these
forms are given in the appendix. Also, note that in this section we do not
explicitly indicate the degree or the ghost number of a form.)
We can now take as the covariant exterior derivative of the $p$-brane wave
functional $\Phi$ the expression
\eq{ \Dcov\Phi=D\Phi + T(\O_{\d x})\Phi +\int(i_{\d x} B + C_{\d x})\Phi.
\eqn{DPHI}}
The left hand side of this equation has to be interpreted as the one-form in
loop space that on an arbitrary vector
$\Psi$ takes the value
\eq{ \Dcov\Phi(\Psi)=\Psi(\Phi) + T(\O_\psi)\Phi +\int(i_\psi B + C_\psi)\Phi,
\eqn{DPHIOFPSI}}
where the symbol $i_\psi$ represents the contraction of the first index of the
$(p+1)$-form $B$ by the background vector
$\psi$, and the pull-back is, as always, understood. It will be clear from the
following calculation why one must add the
integral term in \ref{DPHIOFPSI} to the covariant exterior derivative.
Furthermore, eq.\ \refbr{DPHIOFPSI} corresponds exactly to
the covariant derivative discussed previously in the literature
\cite{BHPSS,BDS91}.

We now want to find the equations (in ordinary spacetime) that the forms
$\O_\psi$ and $C_\psi$ must satisfy in order for
\refbr{DPHI} to be covariant. The BRST variation of \refbr{DPHIOFPSI} is
\eqs{ \d( \Dcov\Phi(\Psi)) &=& \Psi((T(R) + \int\Lambda)\Phi) +
T(\d\O_\psi)\Phi + T(\O_\psi)(T(R) + \int\Lambda)\Phi \lb
&+&\int(\d i_\psi B+ \d C_\psi)\Phi + \int(i_\psi B+ C_\psi)(T(R) +
\int\Lambda)\Phi .\eqn{BRSTONE}}
In order for $\Dcov$ to be covariant, equation \refbr{BRSTONE} must coincide
with
\eq{(T(R) + \int\Lambda)\Dcov\Phi. \eqn{BRSTTWO}}
Taking into account the relations
\eqs{ \Psi(T(R)\Phi) &=& T(\LL_\psi)\Phi + T(R)\Psi(\Phi) \lb
      \Psi(\int\Lambda\Phi) &=& (\int\LL_\psi\Lambda)\Phi +
\int\Lambda\Psi(\Phi) \ =\
      (\int i_\psi d\Lambda)\Phi + \int\Lambda\Psi(\Phi), \lb
      \int(\d i_\psi B)&=& \int(-i_\psi d\Lambda + ni_\psi\o^1_{p+1}),
      \eqn{LEMMAS}}
and the commutator of $T(R)$ with $T(\O_\psi)$ that we have assumed for our
algebra, one can read off the
``covariance equations'' for the background forms in spacetime:
\eqs{\LL_\psi R+\d\O_\psi+[\O_\psi,R]+\tilde{k}[d\O_\psi,dR]&=&0,\lb
     \int\Big(k\,\tr(\O_\psi dR)+ni_\psi\o^1_{p+1}+\d C_\psi
     \Big)&=&0. \eqn{COVARIANCE}}
These two equations express in a compact and coordinate-free notation the
covariance of the operator $\Dcov$.
The solutions for the three-brane and the five-brane are given in the appendix.
Here, we simply want to remark that
for the string case $(p=1)$ one obtains the known loop space results
\cite{BDS91} $R=-w$, $\O_\psi=i_\psi A$ and
$C_\psi = i_\psi A\,A$. Also notice that the variation of $B$ in $\Dcov$ has
cancelled against the Lie derivative
of $\Lambda$, leaving only a $B$-independent term $\o^1_{p+1}$.

We can now immediately construct the curvature tensor associated with the
covariant derivative $\Dcov$.
In the same way as for ordinary Yang--Mills theory, one can define the total
curvature tensor $\GG$ as the
multiplicative operator such that $\Dcov\Dcov\Phi=\GG\Phi$. (Note that this
formula will be modified if loop
space is not torsion-free. This is what happens, for example, in the
supersymmetric formulation of the problem \cite{BHPSS,BDS91}.)

By calculations analogous to those performed when deriving \refbr{COVARIANCE}
one obtains,
\eqs{ \GG(\Psi^1,\Psi^2)&=&T\left(\LL_{\psi^1}\O_{\psi^2}-
     \LL_{\psi^2}\O_{\psi^1}-\O_{[\psi^1,\psi^2]}+
     [\O_{\psi^1},\O_{\psi^2}]+\tilde{k}[d\O_{\psi^1},
     d\O_{\psi^2}]\right) \lb
     &+&\int \left(i_{\psi^1}dC_{\psi^2}-
     i_{\psi^2} dC_{\psi^1}-C_{[\psi^1,\psi^2]} +
     i_{\psi^2} i_{\psi^1} dB +
     k\,\tr\,\O_{\psi^2} d\O_{\psi^1}\right). \eqn{GG}}
In deriving \refbr{GG} we used the facts that $\LL_\psi = i_\psi d + d i_\psi$
and that
     \eq{\int\LL_{\psi^1}i_{\psi^2}B-
     \LL_{\psi^2}i_{\psi^1}B-i_{[\psi^1,\psi^2]}B=\int i_{\psi^2}i_{\psi^1}dB.}
The total curvature $\GG$ depends on both background forms $A$ and $B$. It
is, however, more convenient to split it into the sum of two pieces,
$\GG=\FF+\HH$, where
$\FF$ transforms {\it covariantly} and $\HH$ is {\it invariant} under BRST
transformations.
Furthermore, all the $B$ dependence should be lumped into $\HH$ where it really
belongs.

Naively, one might consider defining $\HH$ simply as
\eq{\HH^0(\Psi^1,\Psi^2)=\int i_{\psi^2} i_{\psi^1} dB . \eqn{HNAIVE}}
However, this is not the right choice, as already has been noticed for the
string. The problem is that $\HH^0$
is not invariant under BRST transformations. Instead, it changes according to
\eq{ \d\HH^0(\Psi^1,\Psi^2)=n\int i_{\psi^2} i_{\psi^1}d\o^1_{p+1}=
      n \d\!\int i_{\psi^2} i_{\psi^1} \o^0_{p+2}.\eqn{DELTAHH0}}
We must therefore add and subtract the Chern-Simons form $\o^0_{p+2}$ to the
total expression $\GG$ and consider the splitting:
\eqs{\FF(\Psi^1,\Psi^2)&=&T\left(\LL_{\psi^1}\O_{\psi^2}-
     \LL_{\psi^2}\O_{\psi^1}-\O_{[\psi^1,\psi^2]}+
     [\O_{\psi^1},\O_{\psi^2}]+\tilde{k}[d\O_{\psi^1},
     d\O_{\psi^2}]\right) \nopagebreak[4]\lb
     &+&\int \left(i_{\psi^1}dC_{\psi^2}-
     i_{\psi^2} dC_{\psi^1}-C_{[\psi^1,\psi^2]} +
     n i_{\psi^2} i_{\psi^1} \o^0_{p+2} +
     k\tr\O_{\psi^2} d\O_{\psi^1}\right), \pagebreak\lb
     \HH(\Psi^1,\Psi^2)&=&\int\left( i_{\psi^2} i_{\psi^1} dB -
     n i_{\psi^2} i_{\psi^1} \o^0_{p+2}\right).\eqn{FPLUSH}}
This construction generalizes a known fact in string theory, namely, that it is
necessary to shift the naive curvature
tensor by the Chern-Simons form $\o^0_3 = \tr(AdA+{2\/3}A^3)$. It is then
straightforward to check that
$\d\HH=0$ and $\d\FF=[T(R)+\int\Lambda, \FF]\equiv[T(R),\FF]$.

The Bianchi identities also follow from our formulation; because of the
manifestly covariant way in which these objects
have been defined in loop space, it is clear that the three-form $\Dcov\GG$
vanishes. Also, because
of the manifestly covariant way in which the splitting \refbr{FPLUSH} has been
performed, it follows that $\Dcov \FF =0$ and $\Dcov \HH\equiv D\HH =0$.

This concludes the construction of the covariant derivative and the related
curvature tensor in loop space. In the next
section, after a review of the concept of (non-abelian) Lie algebra extensions,
we will show how our algebra fits into
this mathematical framework.

\section{General theory of extensions}

In this section we will be reviewing some more mathematically oriented issues
related to Lie algebra extensions, with
particular emphasis on non-abelian extensions. This is the
``infinitesimal version'' of the theory of group extensions
\cite{Eil49,McL50,Kir76}
and it is of interest here because the algebra presented in this
paper becomes a non-abelian extension of the algebra of the charge densities
for $p\geq 5$. The subcase of abelian extensions,
and, in particular, the further subcase of central extensions, are well-known
to physicists due to the abundance of situations
in which they arise. Surprisingly, there has been very little discussion in the
physics literature about the non-abelian case.
Hence, it may be worth presenting a self-contained review of the subject,
paying particular attention to the new aspects that
might be of relevance to physics in the future.
In principle, this section could be read independently from the rest of the
paper, but at the end we will make contact with the
work on the $p$-brane by using our new algebra as an explicit example.

Let us then begin with the definition of an extension of a Lie algebra.
Consider the following exact sequence of Lie algebras,
$0$ denoting the trivial algebra  with one element:
\eq{ 0\stackrel{\d}{\rightarrow} \LLO \stackrel{i}{\rightarrow} \LLh
\stackrel{\pi}{\rightarrow} \LL \stackrel{\e}{\rightarrow} 0.
\eqn{EXACT}}
By exactness of this sequence we mean that all the arrows represent Lie algebra
homomorphisms and that the image of any one
of them coincides with the kernel of the following. In particular, ${\rm
Ker}(i) = {\rm Im}(\d)=0$ means that $\LLO$ is embedded
in $\LLh$ by a one-to-one homomorphism, and $\LL={\rm Ker}(\e) = {\rm Im}(\pi)$
means that $\pi$ is onto. Less trivially, the
condition ${\rm Ker}(\pi) = {\rm Im}(i)$ means that $i(\LLO)$ is an ideal of
$\LLh$ and that $\LL=\LLh/i(\LLO)$.
In the following we will always identify $\LLO$ and $i(\LLO)$, $i$ being an
embedding.

If \refbr{EXACT} is exact, we say that $\LLh$ is an extension of the Lie
algebra
$\LL$ by the Lie algebra $\LLO$. (To avoid confusion, we should mention that
sometimes in the mathematical literature, the
opposite statement is given, with $\LLO$ and $\LL$ interchanged in the above
sentence. We will stick with the convention that
is used in physics.)

The above definition is the most general definition of an extension and we will
use it without making further assumptions about
$\LLO$. When the Lie algebra $\LLO$ is non-abelian one may stress this fact by
calling $\LLh$ a non-abelian extension. In the
specific cases when $\LLO$ is abelian (i.e., has identically vanishing Lie
product: $[\LLO,\LLO]=0$), $\LLh$ is called an
abelian extension. Furthermore, if $\LLO$ is contained in the center of $\LLh$
(i.e., $[\LLO, \LLh]=0$) we refer to $\LLh$ as a
central extension.

It is easy to see why the study of extensions is of interest in quantum
physics. Suppose we have a classical system that yields
an algebra $\LL$ in its canonical formulation. It is well known that going to
the quantum theory, by turning the Poisson brackets
into commutators, one may have to add extra terms to solve ordering
ambiguities, effectively enlarging the algebra to $\LLh$.
These terms, however, are proportional to $\hbar$ and therefore, by power
counting, they must form an ideal $\LLO$ of $\LLh$,
i.e.\ $[\LLO, \LLh] \subseteq \LLO$. This is exactly the statement that $\LLh$
is an
extension of $\LL$ by $\LLO$.

In the above discussion, we assumed the existence of $\LLh$ in order to define
the sequence \refbr{EXACT}. In physics, the
standard situation is that we have the Lie algebra $\LL$ and we want to study
its possible extensions. It should be obvious that
simply knowing $\LLO$ is not enough to uniquely determine $\LLh$. For example,
given any two Lie algebras $\LL$ and $\LLO$ one
could always construct their direct sum, and this is certainly not the most
general case.

We will show that the extra information that is needed is almost entirely
encoded in a homomorphism $ \theta$ from the Lie
algebra $\LL$  to the Lie algebra of the exterior derivations of $\LLO$,
denoted by $\ED$. Recall that a derivation in the Lie
algebra  $\LLO$ is a linear map $D$ from the Lie algebra into itself such that
for any two elements $a$ and $b$,
$D [a,b] = [Da, b] + [a, Db]$. Such derivations form a Lie algebra $D(\LLO)$
and it is easily seen that the adjoint algebra
${Adj}(\LLO)$ (i.e., the algebra generated by the adjoint action $Ad$ of $\LLO$
on itself), is an ideal of  $D(\LLO)$. The
algebra $\ED$ of exterior derivations is then defined as the quotient of the
algebra of all derivations by the adjoint algebra.
In terms of exact sequences one can write this as
\eq{ 0 \rightarrow {Adj(\LLO)}\stackrel{j}{\rightarrow} D(\LLO)
\stackrel{P}{\rightarrow} \ED\rightarrow 0. \eqn{EXACTDERIVATION}}

Even the knowledge of $\theta$ does not uniquely fix the extension, but we will
see towards the end that the freedom left at this
point is very limited; in fact, it is essentially reduced to the choice of a
central element.
The more important question we must ask at this point is if it is always
possible, given $\theta$, to construct an extension.

Thus, given the data $\LL$, $\LLO$ and $\theta$, let us try to construct the
Lie algebra $\LLh$.  First of all, consider the set
of all linear maps $\Psi :\LL \to D(\LLO)$ such that $\Psi_\a \in D(\LLO)$ and
$\theta(\a) = P\circ\Psi_{\a}$ for all $\a$ in $\LL$.
Contrary to $\theta$, which is a homomorphism between $\LL$ and $\ED$, these
functions $\Psi$ may {\it fail} to be Lie algebra
homomorphisms between $\LL$ and $D(\LLO)$. The amount by which the map $\Psi$
fails to be a Lie algebra
homomorphism defines an element of the adjoint algebra. In other words, for any
$\a$ and $\b$ in $\LL$ there is an element
$\chi(\a,\b) \in \LLO$ such that
\eq{ \Psi_\a\circ\Psi_\b - \Psi_\b\circ\Psi_\a=
        \Psi_{[\a,\b]} + Ad_{\chi (\a,\b)}. \eqn{FIRST}}
For a fixed $\Psi$, the map $\chi :\LL \wedge \LL \rightarrow \LLO$ is of
course not unique but it is defined up to a two-cochain
\eq{ \eta : \LL \wedge \LL \rightarrow C(\LLO), \eqn{ETACHAIN}}
$C(\LLO)$ being the center of $\LLO$, i.e., the kernel of the adjoint action
$Ad$.

Let us then take an arbitrary pair $\Psi$ and $\chi$ as above and try to define
the Lie product on the vector space
$\LL+\LLO$ as
\eq{ [(\a; a) , (\b; b)]\equiv([\a,\b] ;
         \chi(\a,\b) + \Psi_\a(b) - \Psi_\b(a) + [a, b]).
      \eqn{TRYLIE}}
(The Lie product will always be denoted by a square bracket; the particular Lie
algebra to which it applies is always clear from its argument.)

It should be emphasized that eq.\ \refbr{TRYLIE} is not just a guess, but in
fact the only possible form for the Lie product
on the pairs. To see this, suppose just for a moment that there {\it is} an
extension $\LLh$. One can then choose any cross
section $\s :\LL \to \LLh$ (i.e., $\pi\circ\s= 1$ ) and uniquely decompose any
element $\lambda \in \LLh$ as
$\lambda=\s(\a)+a\equiv(\a,a)$, where $\a\in\LL$ and $a\in i(\LLO)\equiv\LLO$.
The Lie product in $\LLh$ then ensures that
\eq{[\s(\a)+a,\s(\b)+b]=[\s(\a),\s(\b)]+[\s(\a),
b]-[\s(\b),a]+[a,b].\eqn{HYPOLIE}}
Equations \refbr{TRYLIE} and \refbr{HYPOLIE} are really the same equation if
one sets $\chi(\a,\b)=[\s(\a),\s(\b)]-\s([\a,\b])$ and
$\Psi_\a(b)=[\s(\a), b] $. It is easy to check that $\Psi$ and $\chi$ defined
in this way satisfy all the required properties.
In particular, notice that $[\s(\a), b] $ is  an element of $\LLO$ because
$\LLO$ is an ideal, but it does not necessarily represent
an adjoint action because $\s(\a)$ does not, in general, belong to $\LLO$.

Having convinced ourselves that \refbr{TRYLIE} is the only possible expression
for the Lie product, we must now find under
what circumstances the Jacobi identities are satisfied. This yields the
relation
\eq{ \chi(\a,[\b,\g]) + \Psi_\a(\chi(\b,\g)) +\hbox{cyclic}
    =0 \quad\hbox{ for }\quad \a,\b,\g\in \LL. \eqn{JACOBI}}
At this point we encounter a potential obstruction to the construction of a Lie
product in  $\LL+\LLO$. The relation \refbr{JACOBI}
does not necessarily follow for any pair of functions satisfying \refbr{FIRST}.
What follows from \refbr{FIRST} is the weaker
condition that the quantity
\eq{ \o(\a,\b,\g)=\chi(\a,[\b,\g]) + \Psi_\a(\chi(\b,\g)) +\hbox{cyclic}
\eqn{THREECOCYCLE}}
is a three-cocycle valued in the center of $\LLO$, i.e., $\o:
\LL\wedge\LL\wedge\LL \to C(\LLO)$ such that $\*\o=0$.

Many explanations are in order at this point. First of all, we have to specify
what the coboundary operator $\*$ is. For
completeness, we give its expression on an arbitrary $n$-cochain. The case
$n=3$ is of interest here and the cases $n=1,2$ will
also be needed in a short while. Thus,
\eqs{ & & \*\o_n(\a_1,\cdots,\a_{n+1}) \ =  \lb
      & = & \sum_{1\leq i < j \leq
n+1}(-1)^{i+j}\o_n([\a_i,\a_j],\a_1,\cdots,\a_{i-1},\a_{i+1},\cdots
	\a_{j-1},\a_{j+1},\cdots,\a_{n+1}) \lb
      & + & \sum_{1\leq i \leq
n+1}(-1)^{i+1}\Psi_{\a_i}(\o_n(\a_1,\cdots,\a_{i-1},\a_{i+1},\cdots,\a_{n+1})).
\eqn{GENERALCOBOUNDARY}}

This is the usual definition of a coboundary operator, and we know that
$\*^2=0$. However, there is a further non-trivial check we must
perform. In \refbr{GENERALCOBOUNDARY} we explicitly used the function $\Psi$,
but at the beginning of this section we claimed that
the extension is characterized by $\theta$ alone. We must therefore check that
for any other $\tilde\Psi$ corresponding to the same
$\theta$, eq.\ \refbr{GENERALCOBOUNDARY} gives the same result. This is fairly
obvious if one realizes that two such $\Psi$'s can
only differ by an element of the adjoint algebra of $\LLO$, and that this
element acts trivially on the center of $\LLO$ itself. This
shows that the cohomology of these algebras is specified by $\theta$ alone and
not by the particular choices of $\Psi$ and $\chi$.
The last check to perform is that $\o$ defined in \refbr{THREECOCYCLE} is in
the center of $\LLO$, which can be done by
checking that it has vanishing Lie bracket with an arbitrary element of $\LLO$.

Having checked all these points, we can now go back to \refbr{THREECOCYCLE} and
see where the potential obstruction to constructing
the extension can arise. Suppose that a particular choice of $\Psi$ and $\chi$
satisfying \refbr{FIRST} yields a non-zero $\o$ in
\refbr{THREECOCYCLE}. We may then try to shift $\o$ to zero by using the
arbitrariness that we have in the choice of $\chi$, and let
$\chi \to \chi + \eta$, $\eta$ being an arbitrary two-cochain. Performing these
substitutions in \refbr{THREECOCYCLE}, one can see
that $\o$ is shifted by the coboundary of $\eta$: $\o\to \o + \*\eta$. However,
$\o$ need not be a coboundary, only a cocycle.

We therefore come to the conclusion that the potential obstruction in
constructing $\LLh$ is an element of the third Lie algebra
cohomology group $H^3(\LL,C(\LLO))$. Given the triple $\LL$, $\LLO$ and
$\theta$ one can, by using $\Psi$ and $\chi$, construct
{\it uniquely} an element $[\o]\in H^3(\LL,C(\LLO))$ that does not depend on
the particular choice of $\Psi$ and $\chi$ but only on
$\theta$. The Lie algebra $\LLh$ exists {\it if and only if } $[\o] = 0$. Note
that this obstruction is never present in the theory
of abelian extensions. It will also be absent for the non-abelian extensions we
are considering. Its relevance to physics
is not well-understood at the moment; it seems to suggest the possibility of
having some new form of anomaly, of algebraic nature,
when dealing with systems that admit non-abelian extensions.

Before concluding this general review and applying the results to the algebras
of interest in this paper, we must go back to the
question left unanswered in the beginning of this section of how many
inequivalent extensions there are for a fixed, obstruction-free $\theta$.

First, we need to specify what we mean by equivalent extensions; two extensions
${\LLh}_1$ and ${\LLh}_2$ of $\LL$ by $\LLO$
are said to be equivalent if there is a Lie algebra isomorphism $\Phi: {\LLh}_1
\to {\LLh}_2$ that can be written as
$\Phi(\a,a) = (\a, a+\phi(\a))$ for some $\phi: \LL \to \LLO$. It should be
obvious from the above discussion that two extensions
corresponding to two different $\theta$'s cannot possibly be equivalent.

Suppose now for a moment that the two extensions ${\LLh}_1$ and ${\LLh}_2$ {\it
are} equivalent, i.e., that there is a $\phi$ as above.
Then, the pairs $\Psi^1,\chi^1$ and $\Psi^2,\chi^2$, corresponding to
${\LLh}_1$ and ${\LLh}_2$, are related by
\eqs{ \Psi^1_\a &=& \Psi^2_\a + Ad_{\phi(\a)}, \eqn{ISOPSI}\\
\chi^1(\a,\b) &=& \chi^2(\a,\b) + \Psi^1_\a(\phi(\b)) - \Psi^1_\b(\phi(\a)) +
[\phi(\a,\phi(\b)] - \phi([\a,\b]). \eqn{ISOCHI}}

On the other hand, suppose we are given the two extensions above and we want to
find out if they are isomorphic. Since $\Psi^1$ and
$\Psi^2$ correspond to the same $\theta$, one can always find a map $\phi$ such
that \refbr{ISOPSI} is satisfied, and such a $\phi$
is defined up to a one-cochain $\lambda: \LL \to C(\LLO)$. However, from
equation \refbr{FIRST} we can only infer that the quantity
\eq{ \eta(\a,\b) = -\chi^1(\a,\b) + \chi^2(\a,\b) + \Psi^1_\a(\phi(\b)) -
\Psi^1_\b(\phi(\a)) + [\phi(\a,\phi(\b)] - \phi([\a,\b])
\eqn{TWOCOCYCLE}}
is a two cocycle in $C(\LLO)$, i.e.\ $\*\eta=0$. If we try to use the
arbitrariness in the choice of $\phi$ to shift $\eta$ to zero, we
see that by changing $\phi \to \phi + \lambda$ we can shift $\eta$  only by a
coboundary $\eta \to \eta + \*\lambda$. We then come to
the conclusion that the non-equivalent extensions are labelled by the elements
of the second cohomology group $H^2(\LL,C(\LLO))$.
This is precisely the result familiar in the theory of central extensions.
Hence, having fixed $\theta$ we only have the freedom of
adding a central element. This concludes the review of the general theory.

We can now formulate the construction of our algebra in the language explained
above. When presented in this way it seems to be the
most natural possible extension, given our previous knowledge for the
low-dimensional cases. Thus, let $\S_p$ be a $p$-dimensional
manifold thought of as a space-like section of the $(p+1)$-dimensional
worldvolume spanned by the $p$-brane. The charge densities
$\Tt^a(\s)$ are scalar densities of weight one on $\S_p$. To be specific, we
will think of $a$ as an index in the Lie algebra $u(N)$.
We will work in the dual picture and think of $\LL$ as the algebra of true
scalars $\a:\S_p \to u(N)$.  If $\a\in\LL$, the integral
$\int_{\S_p}\a T$ is well-defined. In the algebra $\LL$ we take as our Lie
product the pointwise Lie product
$[\a,\b](\s)\equiv[\a(\s),\b(\s)]$.

We now want to extend this algebra in a way that preserves the full
diffeomorphism invariance on $\S_p$. This suggests us to take
$\LLO$ to be a formal sum of $u(N)$-valued differential forms of degree higher
than zero, plus, possibly, a central extension.
We also want to retain the same expression for the Lie product that worked in
the cases $p=1,3$, i.e., for the pairs $ a = (X, c_X)$,
$X$ being the formal sum of differential forms and $c_X$ a complex number. We
thus take (wedge product always understood)
\eq{ [(X;c_X),(Y;c_Y)] = \big([X, Y]  + \tilde{k}[dX,dY] ; k\int \tr(X
dY)\big). \eqn{PRODUCTINLLO}}
Note that \refbr{PRODUCTINLLO} also arises in the canonical
formulation of $(2+1)$-dimensional non-linear $\sigma$-models
\cite{RSY84,FR92,Fer92}. This is further evidence that the above
construction may be relevant for other applications.

One could try a general expression of the kind $X = X^1_\mu dx^\mu +{1\/2}
X^2_{\mu\nu}dx^\mu\wedge dx^\nu +\cdots$, but we can
easily see that forms of odd degree cannot appear in $\LLO$ if
\refbr{PRODUCTINLLO} has to satisfy the Jacobi identities.
Hence, our ansatz for $\LLO$ is the algebra of formal sums of differential
forms of even degree $j$, $2\leq j <p$,
$X = X^2+X^4+\cdots$, centrally extended by the complex numbers, with Lie
product \refbr{PRODUCTINLLO}. To avoid misunderstanding,
let us stress that forms of degree higher than $p$, arising from the Lie
product, are all vanishing and that the integral is also
assumed to vanish on all forms of degree not equal to $p$.

Finally, we must give the homomorphism $\theta$. For our purposes, it is
sufficient to give the map $\Psi$:
\eq{\Psi_\a((X;c_X))=\big([\a, X] + \tilde{k}[d\a,
dX]; k\int \tr(\a d X)\big). \eqn{OURPSI}}
In spite of its similarity with the Lie product \refbr{PRODUCTINLLO},
equation \refbr{OURPSI} is not an adjoint action since
$\a\not\in\LLO$.
At this point, one can go back to our general discussion and check that no
obstructions arise in the construction of $\LLh$.

One can now easily see what kind of algebras one gets for different values of
$p$. For $p=1$, $\LLh$ coincides with the Kac--Moody
algebra. Indeed, for $p=1$ the $\tilde{k}$ term is not present at all since the
only forms allowed on $S^1$ are zero- and one-forms.
For $p=2$, and for all even $p$'s in general, the $k$ term is not present,
since it represents the integral of a sum of odd-dimensional
forms. In particular, the case $p=2$ coincides with the algebra arising in the
study on non-linear $\s$-models in $2+1$ dimensions.
For $p=3$ one has the Mickelsson--Faddeev algebra, which is the abelian
extension that appears both in the study of gauge theories
with chiral fermions and in the study of the three-brane.

As long as $p\leq 3$, the algebra $\LLh$ is an abelian extension of the algebra
of charge densities. For higher $p$'s, however, and in
particular for $p=5$, the algebra $\LLO$ becomes non-abelian and, consequently,
the algebra $\LLh$ becomes a non-abelian extension of
the algebra of charge densities. An example of a non-vanishing Lie bracket is
the one between forms of degree two.
It is for these values of $p$ that our algebra differs drastically form those
proposed previously in the literature.

\section{Conclusions and comments}

In this paper we have presented an alternative formulation of the
higher-dimensional loop space operators which is based on a new
algebra for the $p$-brane. The form of this algebra is essentially
fixed by the requirements of linearity, closure, diffeomorphism
invariance and the need to accommodate the extensions
already present  in the $p=1$ and $p=3$ cases.

Some of the previously known material, needed mostly for comparison, was
summarized in section 2.
The explicit form of our algebra and the expression for the BRST
operator were given in section 3. In section 4 we presented
the construction of the covariant derivative and its curvature tensor.
Finally, in section 5, we discussed some more mathematical issues
regarding the theory of extensions.

Throughout the paper we restricted ourselves to a local analysis.
It would be of interest to study also the ``global'' properties of an
algebra of this kind. In particular, it must be possible to give a meaning to
the
exponentiation of the algebra and study what further restrictions
are imposed on the parameters $k$ and $\tilde k$.

As mentioned in the introduction, we do not yet have an action from
which our algebra follows by canonical construction. In principle, such an
action could be constructed by using the method of coadjoint orbits.
It remains to be seen whether this can be done maintaining full
diffeomorphism invariance on the $(p+1)$-dimensional worldvolume.

Finally, the generalization of this construction to superspace should not
present much difficulty.
It is particularly for this case that one expects to make connections with the
dual picture of the
superstring.\\[10mm]
{\Large\bf Acknowledgement}\\[4mm]
We are grateful to J. Hallin for help with the development of Mathematica
software handling differential forms.

\newpage
\appendix
\section{Appendix}
\newcommand{\idx}{i_{\!\!\;\d x}\!}
\newcommand{\Odx}{\O_{\d x}}
\newcommand{\pint}{\int_{\S_p}}

\subsection{Anomaly formulae}
\label{App-formulae}
In this appendix we collect explicit expressions for some standard forms
entering in the
text. The particular expressions given here correspond to the convention of
commuting
exterior derivative $d$ and BRST operator $\d$. However, the conversion to the
opposite case is straightforward.

We define the Chern-Simons form $\o^0_{p+2}$ (up to an exact form) by
\eq{
d\o^0_{p+2}(A) = \tr\br{F^{{p+3\/2}}},
}
$F=dA+A^2$ being the curvature two-form of the Yang--Mills field $A$.
For $p=1,3,5$ we have explicitly
\eqs{
\o^0_3(A) & = & \tr\br{FA-{1\/3}A^3}, \\
\o^0_5(A) & = & \tr\br{F^2A-{1\/2}FA^3+{1\/10}A^5}, \\
\o^0_7(A) & = & \tr\br{F^3A-{2\/5}F^2A^3-{1\/5}FAFA^2+{1\/5}FA^5-{1\/35}A^7}.
}
The BRST variation of $\o^0_{p+2}$, in turn, defines the anomaly $\o^1_{p+1}$
by the descent equation
\eq{
\d\o^0_{p+2} = d\o^1_{p+1}.
}
A compact way of writing $\o^1_{p+1}$ is \cite{Zum85,PSez92}
\eq{
\o^1_{p+1} = \tr\br{d\o\,\f_p(A)},
}
where
\eqs{
\f_1 & = & - A, \\
\f_3 & = & -{1\/2}\br{FA+AF-A^3}, \\
\f_5 & = &
-{1\/3}\br{(F^2A+FAF+AF^2)-{4\/5}(A^3F+FA^3)-{2\/5}(A^2FA+AFA^2)+{3\/5}A^5}.
}

\subsection{The $p$-brane BRST operator}
\label{App-BRST}
As was explained
in section 3, nilpotency of the $p$-brane BRST operator,
defined by
\eq{
\d\F = (T(R) + \pint\L^1_p)\F,
}
requires that $R\equiv R_0+R_2+...+R_{p-1}$ satisfies the equations
\eqs{
\eqn{nilp1}
\d R - R^2 - (dR)^2 = 0 \\
\eqn{nilp2}
n\d \o^1_{p+1} - {1\/2} k\,\tr(dR)^2 = 0
}
Here \refbr{nilp1} must be satisfied at each form level $0,2,...p-1$, whereas
only the level $p+1$ is to be considered in \refbr{nilp2}.

We have solved these equations for the cases $p=1,3,5$ by making proper
ans\"atze
and then using Mathematica software, developed specifically for the task, to
perform the computations. There are no principal difficulties in solving the
equations for higher values of $p$, although the number of terms in the ansatz
grows rapidly with $p$. However, the expressions obtained are not very
enlightening
even for lower $p$'s, and they are given here mostly for the sake of
completeness.

For the string and the three-brane, \refbr{nilp2} gives no conditions on $R$
that do not already follow from \refbr{nilp1}. We can therefore use the
solution
for $R_0$ from the string for the three-brane, and the solutions for $R_0$ and
$R_2$ from the three-brane for the five-brane case. We then find the following
expressions:
\eqs{
R_0 & = & -\o, \\
R_2 & = & \tilde{k}\br{{1\/2} [A,d\o] + (s-{1\/2})\{A,d\o\}}, \\
R_4 & = & \tilde{k}^2\Bigg((-{2\/3}+v) A^3d\o + ({1\/3}+u) A^2d\o\,A + u
A\,d\o\,A^2 + v d\o\,A^3 \lb
      & + & t d\!A A\,d\o + (-1-u+v) A\,d\!A\,d\o + (s+t) d\!A\,d\o\,A \lb
      & + & (-{1\/3}+s^2+t) A\,d\o d\!A + (-u+v) d\o d\!A A + ({2\/3}-s+s^2+t)
d\o\,A\,d\!A\Bigg).
}
However, when determining $R_4$ we found that \refbr{nilp2} fixes one of the
free parameters in the expression obtained by solving  \refbr{nilp1} only. If
one
were to proceed to the seven-brane, the latter solution would be the proper one
to take
over from the five-brane case, since \refbr{nilp2}, which gives the nilpotency
condition for the central term, should be imposed only at form level $p+1$.

Apart from the results above, the nilpotency equations also determine the
central charge
according to
\eq{
k = \left\{ \begin{array}{ll}
		2n, & p=1 \\
		-{n\/\tilde{k}}, & p=3 \\
		{5\/2}{n\/{\tilde{k}^2}}, & p=5 \end{array}
	\right..
}

\subsection{The covariant derivative}
\label{App-cov}
In section 4 the following ansatz for a loop space covariant derivative was
made:
\eq{
\Dcov\F = \br{D + T(\Odx) + \pint(\idx\;\! B_{p+1}+C_{p,\d x})}\F,
}
Here $\Odx\equiv\O_{0,\d x}+\O_{2,\d x}+...+\O_{p-1,\d x}$ and
$C_{p,\d x}$ are the unknown forms to be determined by the covariance equations
\eqs{
\eqn{cov1}
\idx\;\! d\!R + d\idx\;\! R + \d\Odx + [\Odx,R] + \tilde{k}[d\Odx,d\!R] = 0, \\
\eqn{cov2}
k\,\tr\,d\Odx d\!R + n\,d\idx\;\!\o^1_{p+1} + d\d C_{p,\d x} = 0,
}
where the former consists of separate equations at form level $0,2,...,p-1$,
while the latter is to be satisfied at level $p+1$ only.

We have solved these equations for $p=1,3,5$, using the results from the
previous
section for $R$ and $k$. To begin with, we found that
\eq{
C_{p,\d x}(A) = -n\,\tr\br{\idx\;\! A \f_p(A)},
}
where $\f_p(A)$ is the form that enters in the anomaly $\o^1_{p+1}$, and that
we have given explicitly for $p=1,3,5$ in section \ref{App-formulae}. For
$\Odx$ we have not been able
to find similarly nice expressions. Below, we list the solutions corresponding
to the
choice $R_2 = {1\/2}\tilde{k}[A,dw]$,in which case at least $\O_{2,\d x}$ can
be written in a fairly compact way:
\eqs{
\O_{0,\d x} & = &  \idx\;\!A, \\
\O_{2,\d x} & = & \tilde{k}[A,\idx\;\!d\!A - d\idx\;\!A + \idx\;\!(A^2)], \\
\O_{4,\d x} & = & \tilde{k}^2\Bigg((-{1\/4}+t+v)d\!A\,d\idx\;\!d\!A
  + (t+v) d\idx\;\!d\!A\,d\!A			\lb
  & + & (-{1\/6}+v) A^2 d\idx\;\!d\!A
  + ({5\/6}-v)  A\,d\!A\,d\idx\;\!A
  + (-{2\/3}-v) A\,d\!A \idx\;\!d\!A 		\lb
  & + & ({1\/12}+v) A\,d\idx\;\!A\,d\!A
  + ({1\/3}-v) A\,d\idx\;\!d\!A\,A
  + (-{11\/12}+v) A \idx\;\!d\!A\,d\!A		\lb
  & + & (-{1\/4}+v) d\!A\,A\,d\idx\;\!A
  + (-{3\/4}+v) d\!A\,A \idx\;\!d\!A
  + ({1\/4}-t-v) d\!A\,d\!A \idx\;\!A		\lb
  & + & (-{3\/4}+v) d\!A\,d\idx\;\!A\,A
  - {1\/4} d\!A \idx\;\!A\,d\!A
  + ({1\/4}+v) d\!A \idx\;\!d\!A\,A		\lb
  & + & (-{5\/12}+v) d\idx\;\!A\,A\,d\!A
  + (-{1\/6}-v) d\idx\;\!A\,d\!A\,A
  + (-{1\/2}+v) d\idx\;\!d\!A\,A^2 		\lb
  & + & (t+v) \idx\;\!A\,d\!A\,d\!A
  + ({1\/12}+v) \idx\;\!d\!A\,A\,d\!A
  + ({4\/3}-v) \idx\;\!d\!A\,d\!A\,A		\lb
  & + & {1\/2} A^3 d\idx\;\!A - A^3 \idx\;\!d\!A
  + ({1\/6}-v) A^2 d\!A \idx\;\!A
  + {3\/4} A^2 \idx\;\!A\,d\!A + {1\/2} A^2 \idx\;\!d\!A\,A	\lb
  & + & ({2\/3}+v) A\,d\!A\,A \idx\;\!A
  - A\,d\!A \idx\;\!A\,A - A \idx\;\!A\,A\,d\!A - A \idx\;\!A\,d\!A\,A
  - {1\/2} A \idx\;\!d\!A\,A^2 		\lb
  & + & ({3\/4}-v) d\!A\,A^2 \idx\;\!A
  - d\!A\,A \idx\;\!A\,A
  + {3\/4} d\!A \idx\;\!A\,A^2 		\lb
  & - & {1\/2} d\idx\;\!A\,A^3
  + ({1\/12}+v) \idx\;\!A\,A^2 d\!A
  + ({4\/3}-v) \idx\;\!A\,A\,d\!A\,A		\lb
  & + & (-{1\/2}+v) \idx\;\!A\,d\!A\,A^2
  + \idx\;\!d\!A\,A^3 					\lb
  & + & A^4 \idx\;\!A - {3\/2} A^3 \idx\;\!A\,A + A^2 \idx\;\!A\,A^2
  - {3\/2} A \idx\;\! A\,A^3 + \idx\;\!A\,A^4\Bigg).
}
As was the case with $R$, the general solution for $\Odx$ for the three-brane
was found to carry over to the five-brane, although in this case this happened
in a
somewhat less trivial manner.
Finally, note that one of the originally four free parameters in $R_4$ was
fixed by
imposing the covariance conditions.

\end{document}